\documentclass{svjour3}                   
\smartqed  %
\usepackage{epsfig}
\usepackage{graphicx}
\usepackage{lineno}
\usepackage{amsmath}
\usepackage{url}
\journalname{arXiv,}
\newcommand{\myfrac}[2]{\displaystyle \frac{#1}{#2}}

\begin{document}

\title{A simulation of a COVID-19 epidemic based on a deterministic SEIR model}

\author{Jos\'{e} M. Carcione$^{1}$ \and Juan E. Santos$^2$ \\ Claudio Bagaini$^3$ \and Jing Ba$^{4(*)}$}
\institute{
$^1$National Institute of Oceanography and Applied Geophysics - OGS, Trieste, Italy. 
\email{jcarcione@inogs.it}  \\
$^2$Hohai University, China; Buenos Aires University, Argentina; and Purdue University, USA. \\
$^3$Haywards Heath, UK. \\
$^4$Hohai University, Nanjing, China. 
(*) Corresponding author.
}

\date{Submitted: \today \ (first version: April 7)}

\maketitle 

\baselineskip 15pt

\parindent 0.in

{\bf Running title}: Modeling the COVID-19 epidemic.

\vspace{0.5cm}

{\rm Submitted to the special topic}:  
\url{https:/www.frontiersin.org/research-topics}: Coronavirus Disease (COVID-19): Pathophysiology, Epidemiology, Clinical Management and Public Health Response.


%\linenumbers

\begin{abstract}
An epidemic disease caused by a new coronavirus has spread in Northern Italy with a strong contagion rate. We implement 
an SEIR model to compute the infected population and number of casualties of this  epidemic. The example may ideally regard the situation in the Italian Region of Lombardy, where 
the epidemic started on February 24, but by no means attempts to perform a rigorous case study in view of the lack of suitable data and uncertainty of the different parameters, namely, the variation of the degree of home isolation and social distancing as a function of time, the number of initially exposed individuals and infected people, the incubation and infectious periods and the fatality rate.  

First, we perform an analysis of the results of the model, by varying the parameters and initial conditions (in order the epidemic to start, there should be at least one exposed or one infectious human). 
Then, we consider the Lombardy case and calibrate the model with the number of dead individuals to date (May 5, 2020) and 
constraint the parameters on the basis of values reported in the literature.  
The peak occurs at day 37 (March 31) approximately, with a reproduction ratio $R_0$ = 3 initially, 1.36 at day 22 and 0.8 after day 35, indicating different 
degrees of lockdown. The predicted death toll is approximately 15600 casualties, with 2.7 million infected individuals at the end of the epidemic. The incubation 
period providing a better fit of the dead individuals is 4.25 days and the infectious period is 4  days, with a
fatality rate of 0.00144/day [values based on the reported (official) number of casualties]. 
The infection fatality rate (IFR) is 0.57 \%, and 2.37 \% if twice the reported number of casualties is assumed. However, these rates depend on the initially exposed individuals.  
If approximately nine times more individuals are exposed, there are three times more infected people at the end of the epidemic and IFR = 0.47 \%. 
If we relax these constraints and use a wider range of lower and upper bounds for the incubation and infectious periods, we observe that a higher incubation period (13 versus 4.25 days) 
gives the same IFR (0.6 \% versus 0.57 \%), but nine times more exposed individuals in the first case. 
Other choices of the set of parameters also provide a good fit of the data, but some of the results may not be 
realistic. Therefore, an accurate determination of the fatality rate and characteristics of the epidemic is subject to the knowledge of precise bounds of the parameters.

Besides the specific example, the analysis proposed in this work shows how isolation measures, social distancing and knowledge of the diffusion conditions help us to understand the dynamics of the epidemic. Hence, the importance to quantify the process to verify the effectiveness of the lockdown. 
\end{abstract}

\section{Introduction}

The most abundant species in nature are viruses, which are parasites, since they cannot replicate themselves.  
Upon replication, some viruses cause serious infectious diseases in human and/or
animals and are medically, socially, and economically important (Spinney, 2017; Adachi, 2020). One of these species is coronavirus. 
An outbreak of pneumonia caused by a novel coronavirus (COVID-19) began (officially) in February 24, 2020, in Northern Italy,
and the number of the newly reported cases still increase. 
Approximately 29000 casualties are reported in Italy at the time of writing (May 5).
The serious danger COVID-19 poses
is reflected in the high number of cases of transmission to
health-care workers, more than 20 \% in Italy. 
The experience in China showed that the use of relative extreme isolation measures in conjunction
with rapid diagnosis has a strong impact on the dynamics of the epidemic; hence, the importance to understand and quantify the process to verify the effectiveness of the isolation measures (e.g., Chowell et al., 2003). 

There is a long history of mathematical models in epidemiology, going
back to the eighteenth century. Bernoulli (1760) used a mathematical
method to evaluate the effectiveness of the techniques of variolation against
smallpox, with the aim of influencing public health policy. Most of the models are compartmental models, with the population divided into classes and assumptions about the time
rate of transfer from one class to another (Hethcote, 2000; Brauer, 2017). We consider a Susceptible-Exposed-Infectious-Removed (SEIR) model to describe the spread of the virus and compute the number of infected and dead individuals. The SEIR model has many versions and the mathematical treatment can be found, for instance, in Hethcote (2000), Keeling and Rohani (2008) and  Diekmann et al. (2013) among others. The goal is to compute the number of infected, recovered and dead individuals on the basis of the number of contacts, probability of the disease transmission, incubation period, recovery rate and fatality rate. The epidemic disease model predicts a peak of infected and dead individuals per day as a function of time, and assumes that births and natural deaths are balanced, since we are dealing with a very short period of time. The population members solely 
decrease due to the disease dictated by the fatality rate of the disease. The differential 
equations are solved with a forward-Euler scheme. 

\section{Theory and differential equations}

When no vaccine is available, the isolation of diagnosed
infectives and social distancing are the only control measures available.
We consider an SEIR epidemic disease model (e.g., Hethcote, 2000; Al-Showaikh and Twizell, 2004; Keeling and Rohani, 2008; Diekmann et al., 2013). The total (initial) population, $N_0$, is categorized in four classes, namely, susceptible, $S (t)$, exposed, $E (t)$, infected-infectious, $I(t)$ and recovered, $R (t)$, where $t$ is the time variable. 
The governing differential equations are 
\begin{equation} \label{1}
\begin{array}{l}
\dot S = \Lambda - \mu S - \beta S \myfrac{I}{N}  , \\ \\
\dot E = \beta S \myfrac{I}{N} -  (\mu + \epsilon) E  , \\ \\
\dot I = \epsilon E  - (\gamma + \mu + \alpha) I  , \\ \\
\dot R = \gamma I - \mu R , 
\end{array}
\end{equation}
where $N = S + E + I + R \le N_0$ in this case, and a dot above a variable denotes time differentiation. 
Equations (\ref{1}) are 
subject to the initial conditions $S(0)$, $E(0)$, $I(0)$ and $R(0)$.  
The parameters are defined as: 

\begin{verse}

\item
$\Lambda$: Per-capita birth rate.

\item
$\mu$: Per-capita natural death rate.

\item
$\alpha$: Virus induced average fatality rate.

\item
$\beta$: Probability of disease transmission per contact (dimensionless) times 
the number of contacts per unit time. 

\item
$\epsilon$: Rate of progression from exposed to infectious (the reciprocal is the incubation period). 

\item
$\gamma$: Recovery rate of infectious individuals (the reciprocal is the infectious period). 

\end{verse}
having units of (1/T), with T: time.
The scheme is illustrated in Figure 1. The choice $\Lambda = \mu =  0 $ and $\epsilon = \infty$ gives the classical SIR model (e.g., d'Onofrio, 2015), while if $\Lambda$ and $\mu$ are not zero, the model is termed endemic SIR model (e.g., Allen. 2017). However, the SIR model has no latent stage (no exposed individuals) and then it is
inappropriate as a model for diseases with an $\epsilon$ such as that of the COVID-19. 

Let us clarify better the meaning of each quantity. 
$N$ is the total number of live humans in the system at time $t$. $S$ is the number of humans susceptible to be exposed and $E$ is the actual number of exposed individuals (a class in which the disease is latent, they are infected but not infectious); people go from $S$ to $E$ depending on the number of contacts with $I$ individuals, multiplied by the probability of infection ($\beta$) (see Figure 1, where $\beta I/N$ 
is the average number of contacts with infection per unit time of one susceptible). The other processes taking place at time $t$ are: exposed ($E$) become infectious ($I$) with a rate $\epsilon$ and infectious  recover ($R$) with a rate $\gamma$. 
Recovered means individual who do not flow back into the $S$ class, as lifelong
immunity is assumed, but it remains to be seen whether the recovered patients from COVID-19 will develop antibodies and achieve lifelong protection. The reciprocals $\epsilon^{-1}$ and $\gamma^{-1}$ are the average disease incubation and infectious periods, respectively. 

$\Lambda$ is the rate of birth and $\mu$ is the natural rate of death, both per unit time. 
The reciprocal $\mu^{-1}$, interpreted as the normal life expectancy (e.g., 70 yr), refers to average  normal deaths (e.g., natural deaths, by normal flu, accidents, etc), not related to the infectious disease. These quantities describe a  model with vital dynamics (endemic model), which have an inflow of births into the $S$ class at rate $\Lambda$ and deaths 
into the other classes at rates $\mu S$, $\mu I$ and $\mu R$ (see Figure 1).  If $\Lambda = \mu N$, the deaths balance the newborns. The number of live people at time $t$ is $ N (t) = S(t)+E(t)+I(t)+R(t)$, that can be lower or higher than $N_0$ depending on the value of $\Lambda$ and $\mu$. In this case, it is lower than $N_0$. 

One of the key parameters, besides $\beta$, is $\alpha$ that represents the disease-related fatality rate (Chowell et al., 2003; Zhang et al., 2013). In a very fast pandemic, we may assume that there are no births and normal deaths (or they balance and $\Lambda = \mu N$), but deaths due to the fatality rate of the disease. 
This rate is an average, because the model does not take into account the age
(a far higher portion of old people die from the disease than young people), the patients preexisting conditions and the healthcare quality. 

In summary, susceptible persons enter the exposed class
with a rate proportional to $\beta$ and remain there for a
mean incubation period $\epsilon^{-1}$, i.e.,  
those already infected with the disease but not able to
transmit it are in the exposed class and progress to
the infectious class, to recover at the rate $\gamma$ and die at the rate $\alpha$.
It is important to recall that the E class has not the symptoms of the disease, because they are incubating it. They will have symptoms when they pass to class I. Individuals in class I may not have symptoms (asymptomatic), but they are infectious, while those in class E are not. Moreover, individuals in class E can move to R without showing symptoms, but they are infectious when they are in class I.

The dead population as a function of time is $D(t) = N_0 - N(t)$, whereas the curve giving the dead people per unit time is 
\begin{equation} \label{3}
\dot D (t) = - \dot N (t) = - (\dot S + \dot E +\dot I + \dot R) (t). 
\end{equation}

Another equivalent approach is an SEIDR model (e.g., De la Sen et al., 2017; Sameni, 2020), where we have to add  
\begin{equation} \label{1D}
\dot D (t) = \alpha I (t) 
\end{equation}
to equations (\ref{1}). 
In Keeling and Rohani (2008, Section 2.2), $\alpha/(\gamma + \mu) = \rho/(1-\rho)$, where $\rho$ is 
the per capita probability of dying from the infection. 
It can easily be shown that equations (\ref{3}) and (\ref{1D}) are equivalent if births and natural deaths compensate.

\subsection{Reproduction ratio}

The basic reproduction ratio, $R_0$, is the classical epidemiological measure associated
with the reproductive power of the disease. For the SEIR model, it is 
\begin{equation} \label{31}
R_0 = \frac{\beta \epsilon}{(\epsilon+\mu) (\gamma+ \alpha + \mu)} 
\end{equation}
(Diekmann et al, 2013; Zhang et al., 2013).
It gives the average number of secondary cases of infection
generated by an infectious individual. Therefore,  it is used to estimate the growth of the virus outbreak.
$R_0$ provides a threshold for the stability of the disease-free equilibrium point. 
When $R_0 < 1$,  the disease dies out; when $R_0 > 1$, an epidemic occurs. The behaviour of SEIR models as a function of  $R_0$ can be found, for instance, in Al-Sheikh (2012). 

\subsection{Infection and case fatality rates}

The infection fatality rate (IFR) is based on all the population that has been infected, i.e., including the undetected  individuals and asymptomatic.  In terms of the recovery and fatality rates, we have 
\begin{equation} \label{IFR}
\mbox{IFR} \ (\%) = 100 \cdot \frac{D_\infty}{R_\infty + D_\infty}, 
\end{equation}
since the total humans that have been infected is the sum of the recovered and dead individuals, where the subscript refers to the end of the epidemic ($t \rightarrow \infty$). 
It can easily be shown that using the last equation (\ref{1}) and equation (\ref{1D}), we obtain 
\begin{equation} \label{IFR1}
{\rm IFR} \ (\%) = 100 \cdot \frac{\alpha I_\infty}{(\alpha + \gamma) I_\infty - \mu R_\infty } \approx 
100 \cdot \frac{\alpha}{\alpha + \gamma} \approx  
100 \cdot \frac{\alpha}{\gamma} , 
\end{equation}
since the term containing $\mu$ is much smaller, because $\mu \ll \alpha \ll \gamma$, and 
equation (\ref{IFR1}) holds approximately at all times, not only at the end of the epidemic. 
On the other hand, the case fatality rate (CFR) considers the number of deaths related to the diagnosed individuals, and it is always CFR $>$ IFR, since the number of diagnosed individuals is lower than the denominator of equation (\ref{IFR}). The CFR is time dependent and is the usually reported value. 

\section{Numerical algorithm}

We solve the differential equations (\ref{1}) by using a forward Euler finite-difference scheme (e.g., Carcione, 2014), discretizing the time variable as $t = n dt$, where 
$n$ is a natural number and $dt$ is the time step. 
Equations (\ref{1}) and (\ref{3}) become after discretization: 
\begin{equation} \label{4}
\begin{array}{l}
S^{n+1} = S^n + dt \left(\Lambda - \mu S^n - \beta S^n \myfrac{I^n}{N^n} \right)  , \\ \\
E^{n+1} = E^n + dt \left[ \beta S^n \myfrac{I^n}{N^n} -  (\mu + \epsilon) E^n \right] , \\ \\
I^{n+1} = I^n + dt \left[ \epsilon E^n  - (\gamma + \mu + \alpha) I^n \right] , \\ \\
R^{n+1} = R^n + dt \left( \gamma I^n - \mu R^n \right) , \\ \\
\dot D^n = - (\dot S^n + \dot E^n +\dot I^n + \dot R^n) (t) , 
\end{array}
\end{equation}
where $\dot D^n$ is the number of dead people only in the specific day $n$. This algorithm yields positive and bounded solutions [e.g., see Brauer (2017) and Problem 1.42(iv) in Diekmann et al. (2013)], and the system converges to an equilibrium, i.e., $S^n+R^n+D^n = S_\infty+R_\infty+D_\infty  = N_0$ for $t \rightarrow \infty$.  

\section{Results}

Let us consider the following base parameters  as an example to analyze the results by varying some of them. $N_0$ = 10 million, $\alpha$ = 0.006/day, $\beta$ = 0.75/day, $\gamma$ = (1/8)/day, 
$\epsilon$ = (1/3)/day, $\Lambda = \mu N$ (balance of births and natural deaths); and initial conditions: 
$S(0) = N_0-E(0)-1$, $E(0)$ = 20000, $I(0)$ = 1 and $R(0)$ = 0. These data is taken from Chowell et al. (2003, Table 1) for SARS and implies 
an average disease incubation (latent period) of 3 days and an infectious period of 8 days.   
The data correspond to no isolation conditions among individuals and an epidemic situation (high $\beta$,  $R_0 = 5.72 > $ 1). 

The time step of the Euler scheme to solve the discretized equations (\ref{4}) is $dt$ = 0.01 day. 
Figure 2 shows the number of individuals in the different classes (a), and the total number of dead people ($D$) and the number of dead people per specific day ($\dot D$) (b). As can be seen, the peak of dead individuals per day is reached at day 30. The high values in Figure 2b do not consider complete home isolation and social distancing measures (or ``suppression"). 
The maximum number of infected individuals is almost 4 million. According to data from China, around 5 \% of people who tested positive to COVID-19 experience severe symptoms and require admission to an intensive-care unit, almost 200 thousands individuals in this case.
Then, the health system would be completely overwhelmed with very high death rates and inability to provide intensive care. A partial ``mitigation" strategy involving social distancing (home isolation of suspect cases  and social distancing of the elderly) would not be enough, and a severe lockdown is required in order to decrease $R_0$ possibly less than 1 (Ferguson et al., 2020).  

Hereafter, we vary the parameters and plot the infected ($I$) individuals, i.e., excluding 
those who are incubating the disease ($E$).  In order the process to start, there should be at least one exposed or one infectious individual. 
Figure 3 shows the number of infected individuals for $R_0 > 1$ (a) and $R_0 \le 1$ (b), where 
all the other parameters are kept constant unless $\beta$, that takes the value 
\begin{equation} \label{311}
\beta \approx (\gamma+ \alpha) R_0, 
\end{equation}
for $\mu$ much smaller than $\gamma$ and $\alpha$ ($\mu^{-1} \approx$ 83 yr in Italy). 
We recall here that $\beta$ is the probability of transmission times the number of contacts per unit time. Basically, reducing $\beta$ (and $R_0$) the peak decreases in intensity but moves to later times for $R_0$ higher than 1 (Figure 3a), although it is wider. There is a significant reduction in the number of infected individuals for $R_0 \le 1$, meaning that strict home isolation is very effective below a given threshold.  

The effect of the initially exposed individuals are shown in Figure 4 for two sets of values of $R_0$, greater (a) and less (b) than 1. 
Figure 4a indicates that more exposed people does not mainly affects the intensity of the peak, but anticipates the spread of the epidemic, so that the location of the peak is highly dependent on $E(0)$. On the other hand, Figure 4b shows that for $R_0 < 1$, the peak location does not change but its intensity does it significantly, indicating an effective ``suppression" of the epidemic,  with more exposed, more infectious. 
Figure 5 indicates that the incubation period (1/$\epsilon$) has also an impact on the results. If $R_0 > 1$ (5a), increasing the period from 3 to 9 days decreases the maximum number of infected individuals by almost half and delays the spread of the epidemic, but the peak is wider. If $R_0 < 1$, the curves behave similarly, but there are much less infected cases. The initially infectious individuals (from one to ten thousand) has no apparent effect on the results, as can be seen in Figure 6, but this is not the case when we deal with the real case history (see next subsection). 
The effects of the infectious period are shown in Figure 7, where, as expected, increasing this quantity delays the epidemic when $R_0 > 1$. Below $R_0$ = 1, the number of infected individuals decreases substantially. 

Let us assume now that  isolation precautions have been imposed and after day 22 $R_0$ changes from 5.72 to 0.1 [a change of $\beta$ according to equation (\ref{311})],  and consider the same parameters to produce Figure 2. The results are shown in Figure 8, where the peak has moved from day 30 to day 25, with  a significant slowing in
the number of new cases. The total number of dead individuals has decreased, and the number of dead individuals per day at the peak has decreased from 22 K to 13 K, approximately. 
Extreme isolation after imperfect isolation anticipates the process.
Figure 9 shows the results if the isolation measures start two days before, at day 20 instead of day 22. The number of casualties decreased from 220 K to 155 K. 

\subsection{The Lombardy case} 

Next, we attempt to model the COVID-19 epidemic in Lombardy  (Italy), where the data is available at
\url{https://github.com/pcm-dpc/COVID-19}.
The time of writing is day 72 (May 5) and the availability of data allows us 
to perform a relatively reliable fit of the total number casualties from day 1 to date.  
On day 69 (May 2), 329 casualties were reported of which 282 are equally distributed in April, since this number is a late report of the hospitals, corresponding to the whole month of April. 
To predict with high accuracy the behaviour of the epidemic is nearly impossible due to many unknown factors, e.g., the degree of spacial distancing, lack of knowledge of the probability of the disease transmission, characteristics of the disease and parameters of the epidemic.  
Uncertainties are related to the parameter $\beta$ that varies with time, while the others are assumed to lie between certain bounds and also contribute to 
the error. Relative predictions of the trend require an analysis of the data, particularly to define the variation of $\beta$ and $R_0$ with time. We do not assume a specific continuous function, but a general approach should consider a partition into discrete periods,  
[$t_0$, $t_1$], [$t_1$, $t_2$] $\ldots$ [$t_{L-1}$, $\infty$], 
guided by the measures taken by the state and the behaviour of the population.  In this case, $t_0$ = 1 day,  
$t_1$ = 22 day and $t_2$ = 35 day, i.e., $L$ = 3, since after $t_1$ (March 16), home isolation, social distancing and partial Nation lockdown started to be effective, as indicated by an inflection point in the curve of casualties per day (see below), although it is debatable that the Italian government followed the same rules as in Wuhan, China. We also observe that at $t_2$ (March 29), the curve starts to bend downwards and reach a ``peak". 
This partition in three periods is valid to date, but the trend can have an unpredictable behavior due to the factors mentioned above, a too early removal of the lockdown conditions, etc. 

The reported infected people cannot be used for calibration, because these data cannot be trusted. The hospitalization numbers cannot be considered to be representative of the number of infected people and  
it is largely unknown at present the number of asymptomatic, undiagnosed infections. 
However, we are aware that even using the number of casualties is uncertain, since there can be an under-ascertainment of deaths, but the figures cannot vary as much as the error related to the infected individuals.
Hence, the reported number of deceased people could possibly be underestimated due to undeclared cases. This number depends on the country (quality of the health system) and average age of the population, but it is certain that this novel virus is more deadly and spreads more quickly than seasonal flu. Moreover, authorities make a distinction between a death occurred ``with the co-action" of the virus and the death ``caused by" the virus. Indeed, only a small percentage of the casualties were in healthy conditions prior the infection and most of the patients were already affected by other illnesses (eg. diabetes, dementia, cancer, stroke). Therefore, we also consider cases where 100 \% more people have actually died per day, compared to the official figures. 

In order to accomplish the fit, we use the simulated annealing algorithm developed by Goffe et al. (1994). The Fortran code can be found in: 
\url{https://econwpa.ub.uni-muenchen.de/econ-wp/prog/papers/9406/9406001.txt}.
The fit is based on the L$^2$-norm and yields $\alpha$, $\beta_1$ (before $t_0$), $\beta_2$ (after $t_0$), $\beta_3$ (after $t_1$), $\epsilon$, $E(0)$ and $\gamma$ from the beginning of the epidemic (day 1, February 24) to date (day 72, May 5), i.e., seven free parameters. We use the total number of deaths for the calibration. 

Table 1 shows the constraints, initial values and results for different cases, where 
Cases 1 and 2 correspond to approximately nine times less exposed individuals at the beginning of the epidemic, and Cases 2 and 3 assume double casualties. Cases 4 and 5 consider a wider range  of the lower and upper bounds for the incubation and infectious periods ($\epsilon^{-1}$ and $\gamma^{-1}$). The last column do not correspond to variables but indicates the infected individuals at the end of the epidemic, i.e., $I_\infty = R_\infty + D_\infty \approx R_\infty $ and the day of the last infected individual (the end of the epidemic in theory). 
The results are very sensitive to variations of the parameter $\beta$, and consequently those of $R_0$, mostly due to the impact of the performed intervention strategies. 

Figure 10 shows the curves of Case 1 compared to the data (black dots), with IFR = 0.57 \% and $R_0$ decreasing from 
3 to 0.8 at the end of the epidemic. The final infected individuals are 2.69 million people (see Figure 11a and Table 1). The peak value of the I class is 0.3 M or 300 thousand individuals. 
If 5 \% of these people require admission to an intensive-care unit (ICU), it amounts to 15000 individuals and largely
exceeds the capacity of Lombardy, which was approximately 1000 ICU at March 16.  
Figure 11b compares the infectious and dead individuals (per day) and, as expected, the two curves are synchronous, since a proportion $\alpha$ of infectious individuals die. The inflection point at day 22 (Figure 10b) indicates that the isolation measures started to be effective. 
Strict isolation could not be achieved at day 22 due to several 
reasons and there is a reasonable delay of a few days before it can be implemented (day 35).  
The total number of casualties is approximately 15600 and  
the effective duration of the epidemic is about 100 days. However, see the last column indicating the day when the last individual is infected, obtained with the condition $I < 1$. Recent data reveal that the effective duration of the Wuhan epidemic was almost 60 days (Wu et al., 2020, Fig. 1b), a shorter  period favoured by the very strict isolation measures applied in that city. Case 2, that considers twice more casualties, and whose results are shown in Figure 12, has a high fatality rate, IFR = 2.37 \%, but 1.33 million infected people. If the exposed individuals are much higher (Case 3), we obtain IFR = 0.47 \% and 6.5 million infected people (see Figures 13 and 14 and Table 1), but in this case, the fit is not optimal at the beginning of the epidemic. 
The calculations indicate the uncertainty related to the initially exposed individuals, i.e., those that are incubating the disease. 

In the following, we do not show the plots, but the results honour the data. If we modify the constraints and use a wider range of lower and upper bounds, the results are those of Cases 4 and 5 in Table 1. Case 4 has slightly higher incubation and infectious periods 
compared to Case 1, but a higher IFR (2.25 \% versus 0.57 \%), 
whereas more exposed individuals yield an incubation period of 13 days  and lower IFR (Case 5). 
Case 6 considers that initially there are a few exposed individuals (we start with one). The algorithm gives a very good fit of the data with IFR = 3.6 \%, comparable periods to Case 1 and 0.44 M infected individuals. Case 7, that considers $I(0)$ = 1 and starts with one exposed individual, has a good fit, but IFR is too high, possibly wrong, indicating that at day 1 there were more  exposed and infectious individuals. Less initially exposed and infectious individuals requires a higher IFR to fit the curve, but also 
a higher $R_0$ could have the same effect if the IFR is kept within a realistic range. 
Finally, we constraint the incubation and infectious periods between 10 and 20 days, and the results are those of Cases 8 and 9 assuming different initially exposed individuals. The calculations yield fatality rates comparable to that of SARS (Chowell et al., 2003), as Case 6. These calculations indicate the uncertainty in the determination of the parameters of the epidemic, but the solutions has to be restricted to reasonable values of the properties of the disease and parameters of the epidemic. 

The values in Table 1 can be compared to figures reported in the literature. The fatality rate and IFR depend on the age of the population.  
Verity et al. (2020, Table 1) estimate for China an IFR = 0.657 \% but over 60 yr age this rate is 3.28 \%. If the number of infected people is several times higher than the reported cases, the fatality rate could be considerably less than the official one, suggesting that this disease is less deadly than SARS and MERS, although much more contagious. Read et al. (2020) report a mean value $R_0$ = 4, while Wu et al. (2020) obtain values between 1.8 and 2. According to Chowell et al. (2003), IFR = 4.8 \% for SARS, and  
Verity et al. (2020) state that the average case fatality rate (CFR) of SARS is higher than that of COVID-19, with the latter approximately 1.38 \% (their IFR is 0.657 \%).
However, this virus seems to be much more contagious. 
The meaning of $\alpha^{-1}$ is the life expectancy of an individual in the infectious class, i.e, if $\alpha$ = 0.00144/day (Case 1),  the expectancy is 694 days. 

\subsection{Further comments} 

There are more complex versions of the SEIR model as, for instance, including a quarantine class 
and a class of isolated (hospitalized) members (Brauer and Castillo-Chavez, 2012), or 
generalizing the diffusion equations (\ref{1}) with the use of temporal fractional derivatives.  
The replacement of the first-order temporal derivative by a Caputo fractional derivative of non-natural order provides an additional parameter to fit the data (e.g., Caputo et al., 2011; Mainardi, 2010; Chen et al., 2020). Furthermore, the model can be made two-dimensional by including the spatial diffusion of the virus (e.g., Naheed et al., 2014). An alternative to spatial diffusion models are contact networks. 
The actual compartmental network through which the disease spreads is a very important part of epidemic spreading. The model used in this study is a homogeneous approximation to these  network models (Pastor-Satorras and Vespignani, 2001; Montakhab and Manshour, 2012; Pastor-Satorras et al., 2015).

Moreover, the model can be improved by including others classes. De la Sen et al. (2017) propose an SEIADR model, where A are asymptomatic infectious and D are are dead-infective. In other models, recovered can become again susceptible (e.g., Xia et al., 2016)  and, in addition, there are stochastic models (Allen, 2017), although the calibration becomes extremely difficult with the incomplete data provided by the authorities and the high number of parameters to be found. 
At the end of the epidemic, more precise information about the parameters will be available and the complete data can be used to evaluate the development of $\beta$ (and $R_0$) with time. 

The outbreak of a pandemic can have catastrophic consequences, not only from the point of view of the casualties, but also economically. Therefore, it is essential to absolutely avoid it by taking the necessary measures at the right time, something that has not been accomplished in Italy and the rest of the world.  According to these calculations, the effective measures are  social distancing and home isolation, since there is no health system designed for ordinary circumstances that can be prepared for a pandemic, when the infected individuals grows exponentially. As can be seen, the pandemic can develop in a few days and the number of casualties can be extremely high if the fatality rate and contagiousness of the disease are  high. Only a few days to take action can make a big difference in the prevention of this disaster. The pandemic and its consequences have been predicted in October 2019 by a group of experts 
(\url{https://www.politico.com/news/magazine/2020/03/07/coronavirus-epidemic-prediction-policy-advice-121172}), but states ignored the fact and transnational nature of the threat, delaying the necessary measures to avoid the disaster, minimizing in many cases the downsides to their own populations and economies. Moreover, in less than three weeks, the virus has overloaded the health-care system over northern Italy, in particular Lombardy, where the 
system cannot support this type of emergency and the authorities are not prepared to deal with the epidemic. 

\section{Conclusions}

A high number of secondary infections by COVID-19 can take place when an infected individual is  introduced into a community. It is essential to simulate the process of infection (and death) in advance, to apply adequate control measures and mitigate the risk of virus diffusion. One of the most used mathematical algorithms to describe the diffusion of an epidemic disease is the SEIR model, that we have applied to compute the number of infected, recovered and dead individuals on the basis of the number of contacts, probability of the disease transmission, incubation and infectious periods, and disease fatality rate. 

A first analysis of the results of the model is based on parameters of the SARS disease and we assume that 
the parameters do not change during the whole epidemic. Reducing the number of contacts, the peak decreases in intensity but moves to later times, although it is wider. Moreover, more exposed people does not affect the intensity of the peak, but anticipates the epidemic. 
The incubation period has also an impact on the results, with higher values delaying the epidemic. The dependence on the initial number of infected people is weak apparently if $R_0$ does not change during the epidemic. 
Increasing the infectious period has the opposite effect of increasing the incubation period.
Moreover, the day when the isolation starts is important, since only two days makes a big difference in the number of casualties. 

The Lombardy modeling assumes ten million of individuals and has been calibrated on the basis of the total number of casualties. The results show that the peak occurs after 37 days with a final number of dead individuals depending on the reproduction ratio $R_0$. With the present available data, this number is approximately 15600. Up to day 72 (May 5, the day of writing), the reproduction ratio is 3 before March 16 (day 22), 1.36 [between March 16 and March 29 (day 35)] and 0.8 after March 29, whereas the fatality rate is 0.00144/day (IFR = 0.57 \%). 
We have also doubled the number of casualties and obtained 
IFR = 2.37 \% and 0.47 \%, with the second value corresponding to nine times more exposed individuals. These values are obtained by constraining the incubation and infectious periods to values reported in the literature. If we relax these constraints and use a wider range of lower and upper bounds, we obtain slightly higher incubation and infectious periods compared to  
the first case, but a much higher IFR (2.25 \% versus 0.57 \%), while much more exposed individuals yields an incubation period of 13 days and a lower IFR (0.6 \%). Of the many solutions that honour the data, we suggest to 
consider as more realistic ones those which agree with the experimental data published at present, based on the reported incubation and infection periods and IFR. The uncertainty is due to the novelty of the virus, whose properties were unknown two months ago, and to the initial conditions, i.e., the initially exposed and infectious individuals.

The present data fit and consequent prediction of the epidemic does not take into account the second phase  established by the state, which started on May 4. After the partial opening of the economy and under a less stringent lockdown, the reproduction number could increase and induce a second outbreak of the epidemic. 
Therefore, a precise determination of the fatality rate is subject to the knowledge of the parameters of the epidemic and characteristics of the disease, and it is clear from these calculations that 
the usefulness of simple models to predict is limited, and that their main role is to help in our
understanding of the dynamics of the epidemic. 

Models can be used to predict and understand how an infectious disease spreads in the 
world and how various factors affect the dynamics. Even if the predictions are 
inaccurate, it has been clear to scientists from many decades to date that quarantine, social distancing and 
the adoption of very strict health and safety standards are essential to stop the spreading of the virus. Quarantine  was even implemented in medieval times to fight the black death before knowing the existence of viruses. 
In this sense, this pandemic reveals the failure of policy makers, since it is well known from basic modeling results that anticipating those measures can save thousand of lives and even prevent the pandemic. The interface of science, society and politics is still uneasy, even in highly developed countries, revealing
a disregard for scientific evidence. Moreover, one of the consequences is that some of these countries do 
not invest sufficiently in R\&D and must acquire the new technology overseas at a much higher cost. 

\vspace{1cm}

{\bf Author contributions}: JMC has written the theory and made the analysis; JES verified the figures and contributed to the analysis; CB aided in the solution of the differential equations; JB contributed to the discussion. 

\vspace{1cm}

{\bf Conflict of interest}:  The authors declare that the research was conducted in the absence of any commercial or financial relationships that could be construed as a potential conflict of interest.

\vspace{1cm}

{\bf Acknowledgements}: JMC dedicates this work to Antonio Buonomo, who passed away of COVID-19 on March 21, 2020.

\newpage
\vspace{5mm}

{\bf References} 

\vspace{-1cm} 

\begin{verse}

\item
Al-Sheikh, S. (2012). 
Modeling and analysis of an SEIR epidemic nodel with a
limited resource for treatment, 
{\it Global Journal of Science Frontier Research, 
Mathematics and Decision Sciences}, 
Volume 12 Issue 14. 

\item
Adachi, A. (2019).  
Grand challenge in human/animal virology: Unseen, smallest replicative entities shape the whole globe, {\it Frontiers in Microbiology}, 11, article 431. 

\item
Allen, L. J. S. (2017). 
A primer on stochastic epidemic models: Formulation,
numerical simulation, and analysis, 
{\it Infectious Disease Modelling}, 2(2), 128--142.  

\item
Al-Showaikh, F., and Twizell, E. (2004). One-dimensional measles dynamics, {\it Appl. Math. Comput.}, 152, 169--194.

\item
Bernoulli, D. (1760). Essai d'une nouvelle analyse de la mortalit\'e caus\'ee par la petite v\'erole et des avantages de l'inoculation pour la pr\'evenir, {\it M\'emoires de Math\'ematiques et de Physique, Acad\'emie Royale des Sciences}, Paris, 1--45.

\item
Brauer, F. (2017). Mathematical epidemiology: Past, present, and future, 
{\it Infectious Disease Modelling}, 2(2), 113--127. 

\item
Brauer, F., Castillo-Chavez, C. (2012). {\it Mathematical Models in
Population Biology and Epidemiology}. Springer, New York.

\item
Caputo, M., Carcione, J. M., and Cavallini, F. (2011). Wave simulation in biological
media based on the Kelvin-Voigt fractional-derivative stress-strain relation, {\it Ultrasound
in Med. \& Biol.}, 37, 996--1004.

\item
Carcione, J. M. (2014). {\it Wave Fields in Real Media. Theory and numerical simulation 
of wave propagation  in anisotropic, anelastic, porous and electromagnetic media}, 3rd edition, Elsevier.  

\item
Chen, Y., Cheng, J., Jiang, X., and Xu, X. (2020). 
The reconstruction and prediction algorithm of the fractional
TDD for the local outbreak of COVID-19,  
\url{https://arxiv.org/abs/2002.10302}

\item
Chitnis, N., Mac Hyman, J., and Cushing, J. M. (2008). 
Determining important parameters in the spread of malaria through the sensitivity analysis of a mathematical model, {\it Bulletin of Mathematical Biology}, 70(5), 1272--1296. 

\item
Chowell, G., Fenimore, P. W., Castillo-Garsow, M. A., Castillo-Chavez, C. (2003). SARS outbreak in Ontario, Hong Kong and Singapore: the role of diagnosis and
isolation as a control mechanism, {\it J. Theor. Biol.}, 224, 1--8.

\item
Diekmann, O., Heesterbeek, H., and Britton, T. (2013). {\it Mathematical tools for
understanding infectious disease dynamics}.
Princeton Series in Theoretical and Computational Biology. Princeton University
Press, Princeton.

\item
De la Sen, M., Ibeas, A.,  Alonso-Quesada, S., and Nistal, R. (2017). 
On a new epidemic model with asymptomatic and
dead-infective subpopulations with feedback controls
useful for Ebola disease, {\it Discrete Dynamics in Nature and Society}, 
https://doi.org/10.1155/2017/4232971

\item
d'Onofrio, A., Manfredi, P., and Salinelli, E. (2015). 
Dynamic behaviour of a discrete-time SIR model with
information dependent vaccine uptake, 
Journal of Difference Equations and Applications,
\url{http://dx.doi.org/10.1080/10236198.2015.1107549}.

\item 
Ferguson, N. M., et al. (2020). Impact of non-pharmaceutical interventions (NPIs) to reduce COVID-19 mortality and healthcare demand. \url{https://doi.org/10.25561/77482}
         
\item
Goffe, W. L., Ferrier, G. D., and Rogers, J. (1994). Global optimization of statistical functions with simulated annealing, {\it Journal of Econometrics}, 60(1-2), 65--99. 
                       
\item
Hethcote, H. W. (2000). The mathematics of infectious diseases, {\it SIAM Review}, 42,
599--653.

\item
Keeling, M. J., and Rohani, P. (2008). {\it Modeling infectious diseases in humans and animals}. Princeton University Press. 

\item
Lauer, S. A., Grantz, K. H., Bi, Q., Jones, F. K., Zheng, Q., Meredith, H. R., Azman, A. S., Reich, N. G., and Lessler,  J. (2020),  The incubation period of coronavirus disease 2019 (COVID-19) from
publicly reported confirmed cases: Estimation and application, {\it Annals of Internal Medicine}, 
DOI: 10.7326/M20-0504.

\item
Mainardi, F. (2010). {\it Fractional Calculus and Waves in Linear Viscoelasticity}, Imperial College Press, London.

\item
Montakhab, A., and Manshour, P. (2012). 
Low prevalence, quasi-stationarity and power-law behavior in a model of contagion spreading, 
{\it Exploring the Frontiers of Physics}, 99, 58002-p1--6.

\item
Naheed, A., Singh, M., and Lucy, D. (2014). Numerical study of SARS epidemic model
with the inclusion of diffusion in the system, {\it Applied Mathematics and Computation}, 229, 480--498.

\item
Pastor-Satorras, R., and Vespignani, A. (2001). 
Epidemic spreading in scale-free networks, {\it Phys. Rev. Lett.}, 86, 3200--3203. 

\item
Pastor-Satorras, R., Castellano, C., Van Mieghem, P., and Vespignani, A. (2015).  
Epidemic spreading in complex networks, {\it Rev. of Modern Physics}, 87, 925--979. 

\item
Read, J. M., Bridgen, J. R. E., Cummings, D. A. T., Ho, A., Jewell, C. P. (2020).  
Novel coronavirus 2019-nCoV: early estimation of epidemiological parameters and epidemic predictions, 
https://doi.org/10.1101/2020.01.23.20018549. 

\item
Sameni, R. (2020). 
Mathematical modeling of epidemic diseases; A
case study of the COVID-19 coronavirus, 
\url{https://arxiv.org/abs/2003.11371}

\item
Spinney, L. (2017). 
{\it Pale rider: The spanish flu of 1918 and how it changed the world}. Jonathan Cape, London.

\item
Verity, R. et al. (2020).  
Estimates of the severity of coronavirus disease 2019:
a model-based analysis, 
DOI:https://doi.org/10.1016/S1473-3099(20)30243-7.

\item
Xia, W., Kundu, S., and Maitra, S. (2018). Dynamics of a delayed SEIQ epidemic
Model, {\it Advances in Difference Equations}, 336, \url{https://doi.org/10.1186/s13662-018-1791-8}

\item
Wu, J. T., Leung, K., Bushman, M., Kishore, N., Niehus, N., 
de Salazar, P. M.,  Cowling, B. J.,  Lipsitch, M., and Leung, G. M. (2020). Estimating clinical severity of COVID-19 from the transmission dynamics in Wuhan, China, {\it Nature Medicine Letters}, \url{https://doi.org/10.1038/s41591-020-0822-7}

\item
Zhang, L.-J., Li, Y., Ren, Q., and Huo, Z. (2013).  
Global dynamics of an SEIRS epidemic model with constant
immigration and immunity, 
{\it WSEAS Transactions on Mathematics}, 12, 630--640. 

\end{verse}

\newpage

\underline{Manuscript contribution to the field}: We implement an SEIR model to evaluate the date of the infection peak of the COVID-19 epidemic, that includes the disease fatality rate to estimate the number of casualties per day. To our knowledge, this is the first time that this model is calibrated with the number of casualties. The simulation attempts to provide a simple procedure  to model the coronavirus diffusion in a given region. The example regards the epidemic in the Lombardy province (Italy) which is taking place at the time of this writing, but can be applied in general. We show how the date of the peak and the number of casualties are affected by the effectiveness of the home isolation, incubation and infectious periods, probability of transmission and initially exposed individuals. The results from the procedure, that intends to be the basis for a real case study, where it is clear the effect of each parameter and variable in the dynamic evolution of the epidemic. 

\newpage

\begin{center}
\baselineskip 10pt
Table 1. Constraints and initial--final values of the inversion algorithm. 
Several cases that honour the data.
\end{center}
\[
\begin{tabular}{|c|c|c|c|c|c|c|c|c||c|}
\hline
  Case & Variable $\rightarrow$  & $\alpha$ & $\beta_1$ & $\beta_2$ & $\beta_3$ & $\epsilon^{-1}$ & $\gamma^{-1}$ & $E(0)$ & $I_\infty$ (M) \\
 &  & (day$^{-1}$) & (day$^{-1}$) & (day$^{-1}$)  & (day$^{-1}$)   
  & (day) & (day)  & & $L$ (day) \\
 &  &  &  &  & &  &  & & $D_\infty$ \\
\hline
\hline
 & Lower bound   & 10$^{-5}$ & 0.5 & 10$^{-6}$ & 10$^{-6}$ 
& 3 & 3 & 10$^3$ & \\
 & Upper bound   & 10$^{-1}$ & 0.9 & 10$^{3}$ & 10$^{3}$ 
& 6 & 6 & 2 $\times$ 10$^5$  & \\
 & Initial value  & 0.006 & 0.75 & 0.5 & 0.5
& 5 & 5 & 10$^4$/10$^5$ & \\
\hline
1.1 & Final value  & 0.00144 & 0.75 & 0.34 & 0.2 
& 4.25 & 4.02 & 11460 & 2.69 \\
1.2 & IFR & 0.57 \% & & & & &  & & 262--Nov 11 \\
1.3 & $R_0$ & & 3.00 & 1.36 & 0.80 & &  & & 15652 \\
\hline
2.1 & Final value ${(\ast)}$ & 0.0051 & 0.702 & 0.29 & 0.132 
& 5.45 & 4.75 & 9460 & 1.33 \\
2.2 & IFR &  2.37 \% & & & & &  & & 264--Nov 13 \\
2.3 & $R_0$ & & 3.25 & 1.34 & 0.61 & &  & & 31934 \\
\hline
3.1 & Final value ${(\ast)}$ & 0.00142 & 0.75 & 0.57 & 0.395 
& 5.79 & 3.31 & 99500 & 6.49 \\
3.2 & IFR &  0.47 \% & & & & &  & & 236--Oct 16 \\
3.3 & $R_0$ & & 2.47 & 1.87 & 1.30 & &  & & 30544 \\
\hline
\hline
 & Lower bound  & 10$^{-5}$ & 0.5 & 10$^{-6}$ & 10$^{-6}$ 
& 2 & 2 & 10$^3$ & \\
 & Upper bound  & 10$^{-1}$ & 0.9 & 10$^{3}$ & 10$^{3}$ 
& 20 & 20 & 2 $\times$ 10$^5$ & \\
 & Initial value  & 0.006 & 0.75 & 0.5 & 0.2
& 5/15 & 5/15 & 10$^3$/10$^5$ & \\
\hline
4.1 & Final value  & 0.00436 & 0.59 & 0.29 & 0.094 
& 6.10 & 5.28 & 8800 & 0.69 \\
4.2 & IFR & 2.25 \% & & & & &  & & 239--Oct 19 \\
4.3 & $R_0$ & & 3.04 & 1.50 & 0.48 & &  & & 15652 \\
\hline
5.1 & Final value  & 0.0011 & 0.81 & 0.33 & 0.03 
& 13 & 5.53 & 91900 & 2.49 \\
5.2 & IFR  & 0.60 \% & & & & &  & & 247--Oct 27 \\
5.3 & $R_0$  & & 4.45 & 1.81 & 0.16 & &  & & 15345 \\
\hline
\hline
6.1 & Final value  & 0.0073 & 0.755 & 0.23 & 0.125 
& 4.87 & 5.11 & 170 & 0.44 \\
6.2 & IFR  & 3.59 \% & & & & &  & & 269--Nov 18 \\
6.3 & $R_0$  & & 3.72 & 1.13 & 0.61 & &  & & 16401  \\
\hline
7.1 & Final value  & 0.09 & 0.9 & 0.28 & 0.175 
& 2.99 & 6.15 & 32 & 0.04 \\
7.2 & IFR  & 35.62 \% & & & & &  & & 212--Sep 22 \\
7.3 & $R_0$  & & 3.56 & 1.11 & 0.69 & &  & & 16112 \\
\hline
\hline

\hline
8.1 & Final value  & 0.00674 & 0.83 & 0.006 & 0.01 
& 12.79 & 14.93 & 1270 & 0.18 \\
8.2 & IFR  & 9.15 \% & & & & &  & & 272--Nov 21 \\
8.3 & $R_0$  & & 11.2 & 0.08 & 0.13 & &  & & 16681 \\
\hline

\hline
9.1 & Final value  & 0.0055 & 0.506 & 0.044 & 0.01 
& 11.08 & 14.97 & 8960 & 0.22 \\
9.2 & IFR  & 7.60 \% & & & & &  & & 268--Nov 17 \\
9.3 & $R_0$  & & 7.00 & 0.61 & 0.14 & &  & & 16653 \\
\hline

\end{tabular}
\]
%\begin{center}

\small
\footnotesize
%\scriptsize

\baselineskip 10pt
\hspace{0.1cm} 
$I(0)$ = 1000 unless Case 7 with $I(0)$ = 1. 

\hspace{0.1cm} ${(\ast)}$ Doubling the number of casualties. 

\hspace{0.1cm} The values of $\beta$ refer to the periods (in days): [1, 22], [22, 35] and [35, $\infty$] (in days). 

\hspace{0.1cm} $I_\infty$ (in millions) indicates the total infected individuals at the end of the epidemic.

\hspace{0.1cm} $L$ denotes the day of the last infected individual, obtained when $I < 1$.

\hspace{0.1cm} $D_\infty$ is the death toll at the end of the epidemic.

\hspace{0.1cm} Read et al. (2020) report the mean values $\epsilon^{-1}$ = 4 days and $\gamma^{-1}$ = 3.6 days.

\hspace{0.1cm} Lauer et al. (2020) report $\epsilon^{-1}$ = 5.1 days.

\hspace{0.1cm} Ferguson et al. (2020) estimate an average IFR = 0.9 \%.

\baselineskip 23pt
%\end{center}

\newpage

\begin{figure}
\vspace{1cm}
\includegraphics[scale=0.6]{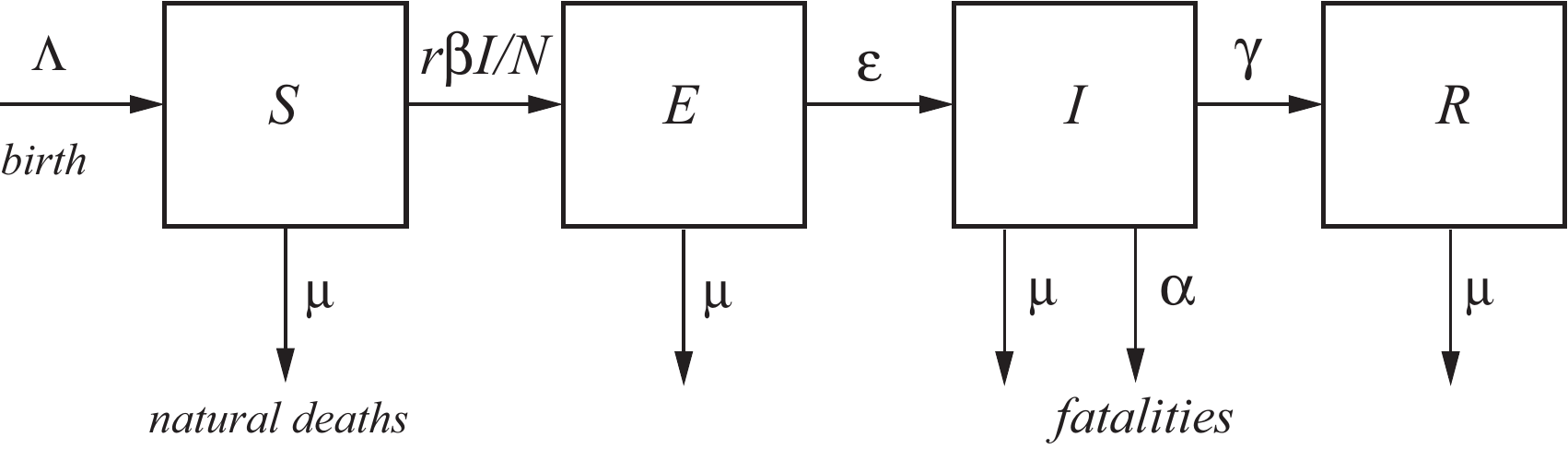}
\caption{A typical SEIR model. 
The total population, $N$, is categorized in four classes, namely, susceptible, $S$, exposed $E$, infected $I$ and recovered $R$ (e.g., Chitnis et al., 2008). $\Lambda$ and $\mu$  correspond to births and natural deaths independent of the disease, and $\alpha$ is the fatality rate.  
}
\vspace{1cm}
\end{figure}

\begin{figure}
\includegraphics[scale=0.7]{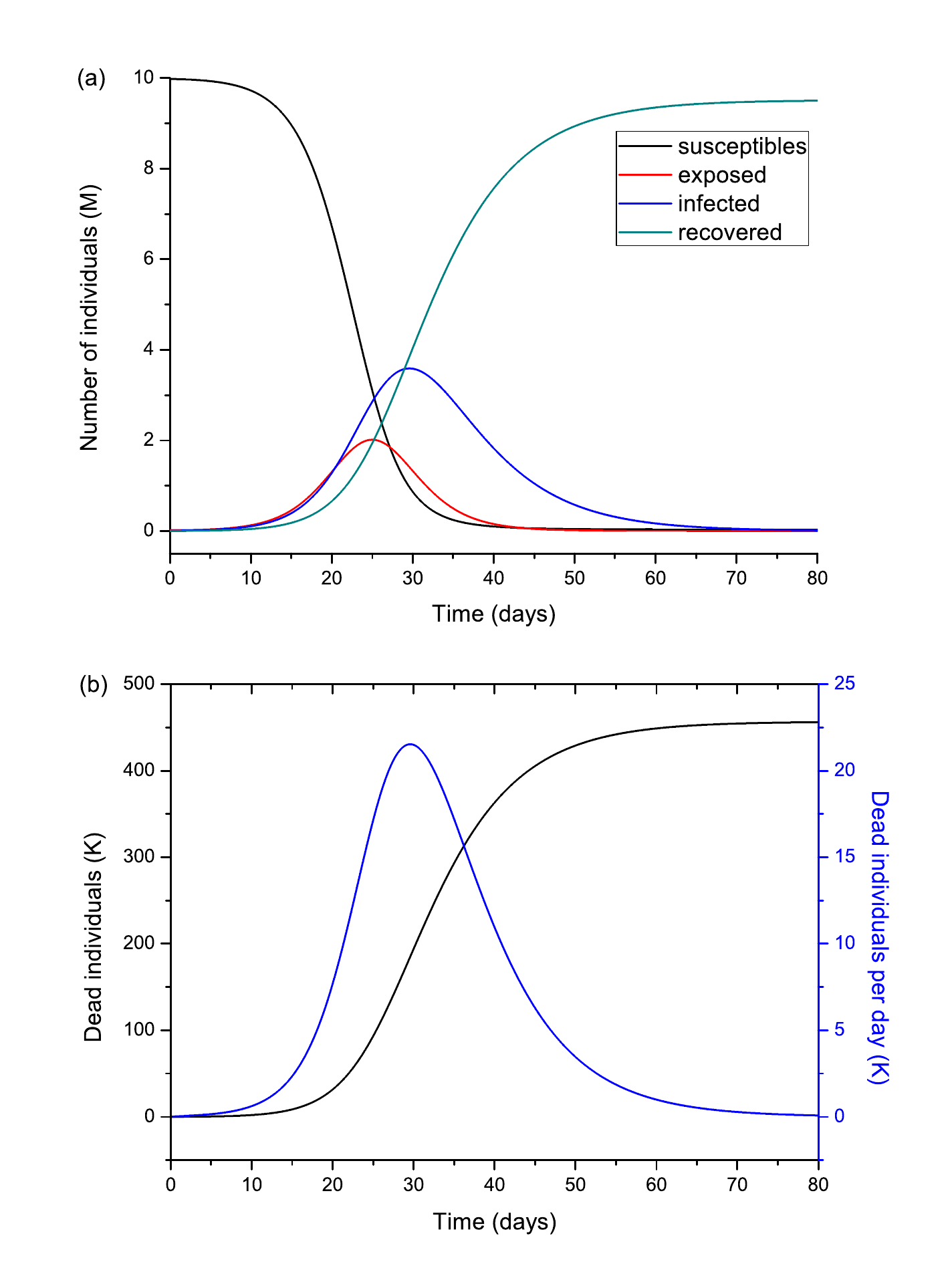}
\caption{Number of humans in the different classes (millions) (a), and total number of deaths and the number of deaths per specific day (thousands) (b). The number of exposed people  at $t$ = 0 is 20000 and there is one initial infected individual, $I(0)$ = 1. The value of $R_0$ = 5.72 means imperfect isolation measures. 
}
\vspace{1cm}
\end{figure}

\begin{figure}
\includegraphics[scale=0.7]{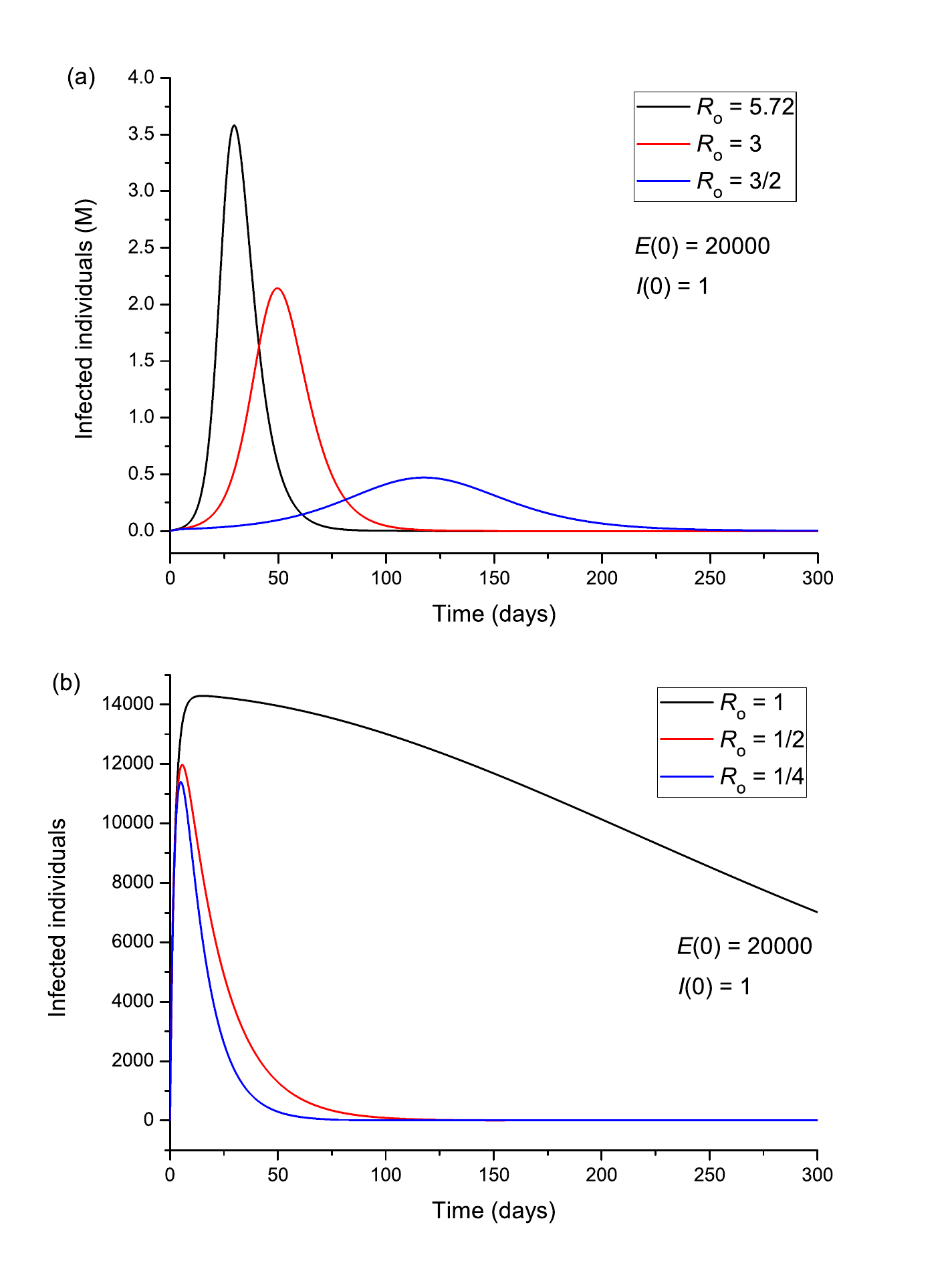}
\caption{Infected individuals for different values of $R_0$, corresponding to values greater (a) and smaller (b) than 1.}
\vspace{1cm}
\end{figure}

\begin{figure}
\includegraphics[scale=0.7]{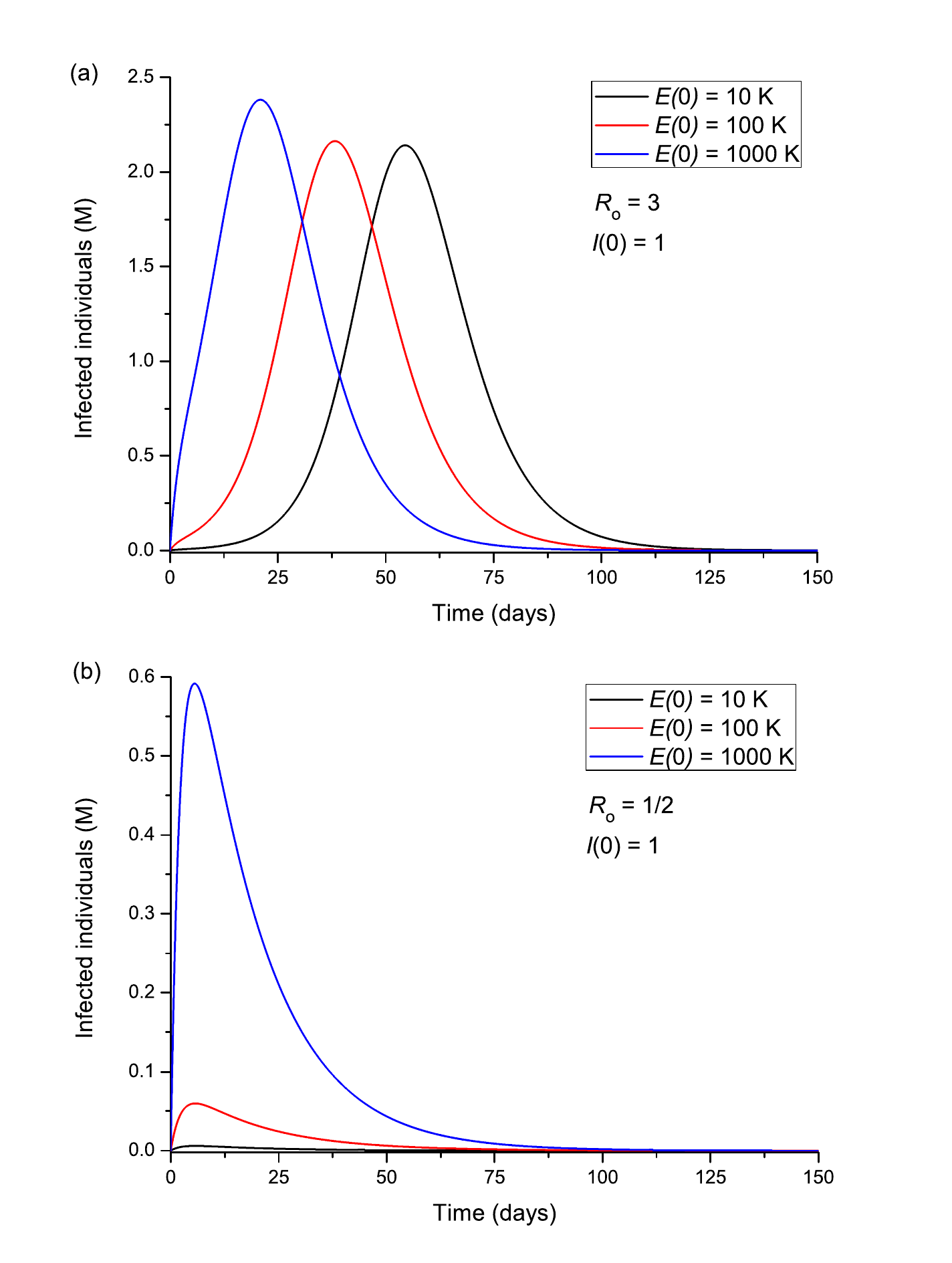}
\caption{Infected individuals for different values of the initially exposed individuals, 
corresponding to $R_0$ greater (a) and smaller (b) than 1.}
\vspace{1cm}
\end{figure}

\begin{figure}
\includegraphics[scale=0.7]{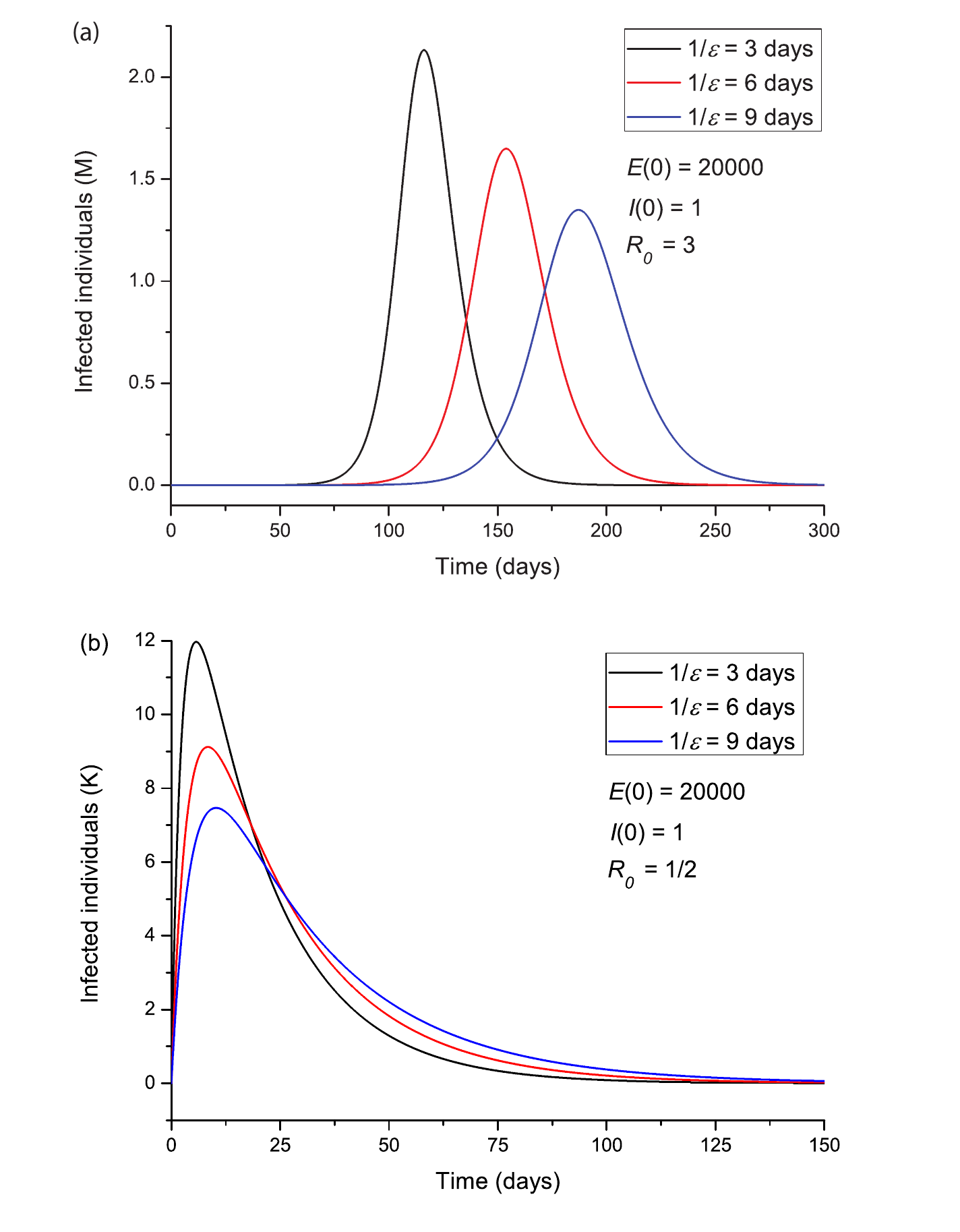}
\caption{
Infected individuals for different values of the incubation period $\epsilon^{-1}$, 
corresponding to $R_0$ greater (a) and smaller (b) than 1. 
}
\vspace{1cm}
\end{figure}

\begin{figure}
\includegraphics[scale=0.7]{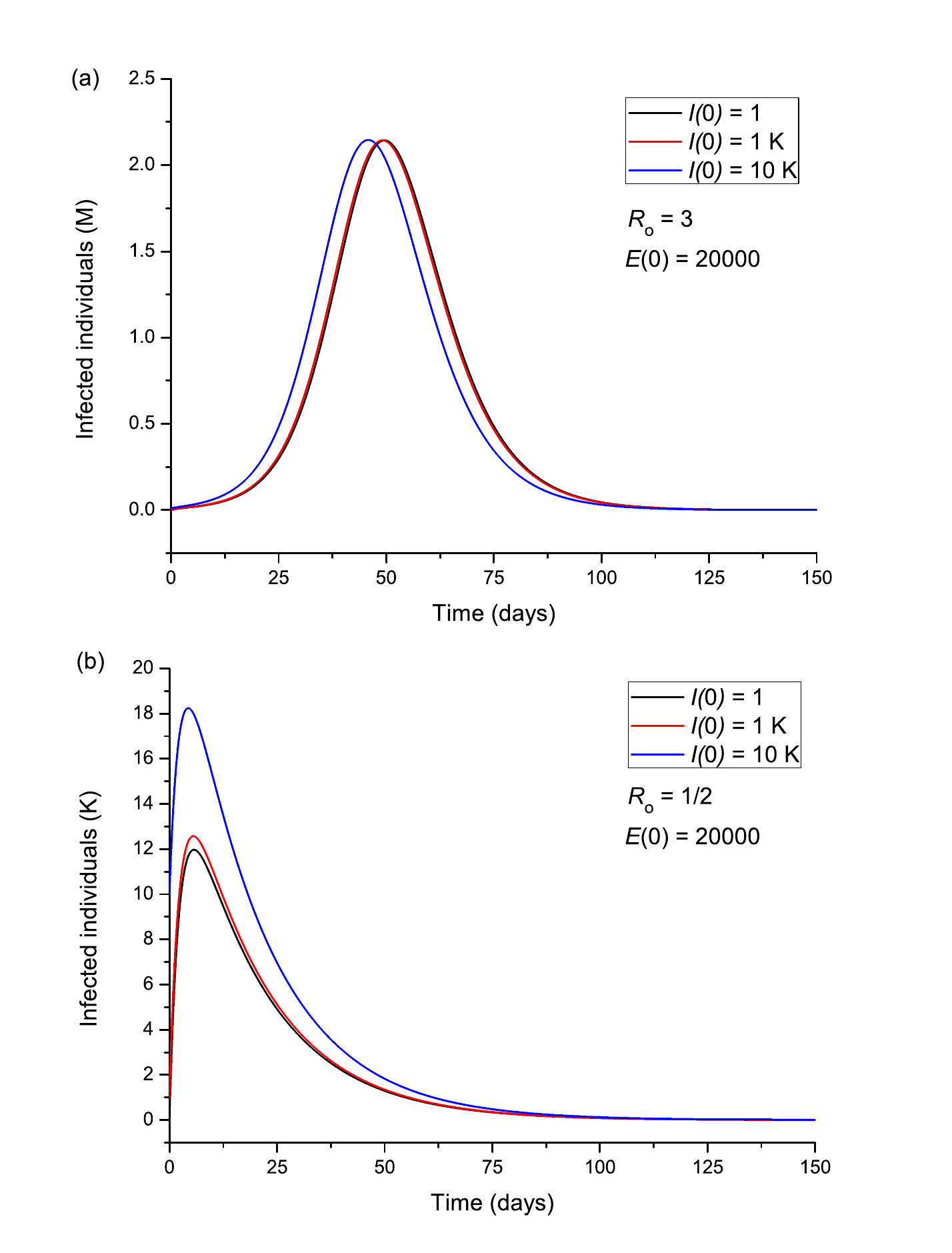}
\caption{Infected individuals for different values of the initially infected individuals, 
corresponding to $R_0$ greater (a) and smaller (b) than 1. 
}
\vspace{1cm}
\end{figure}

\begin{figure}
\includegraphics[scale=0.7]{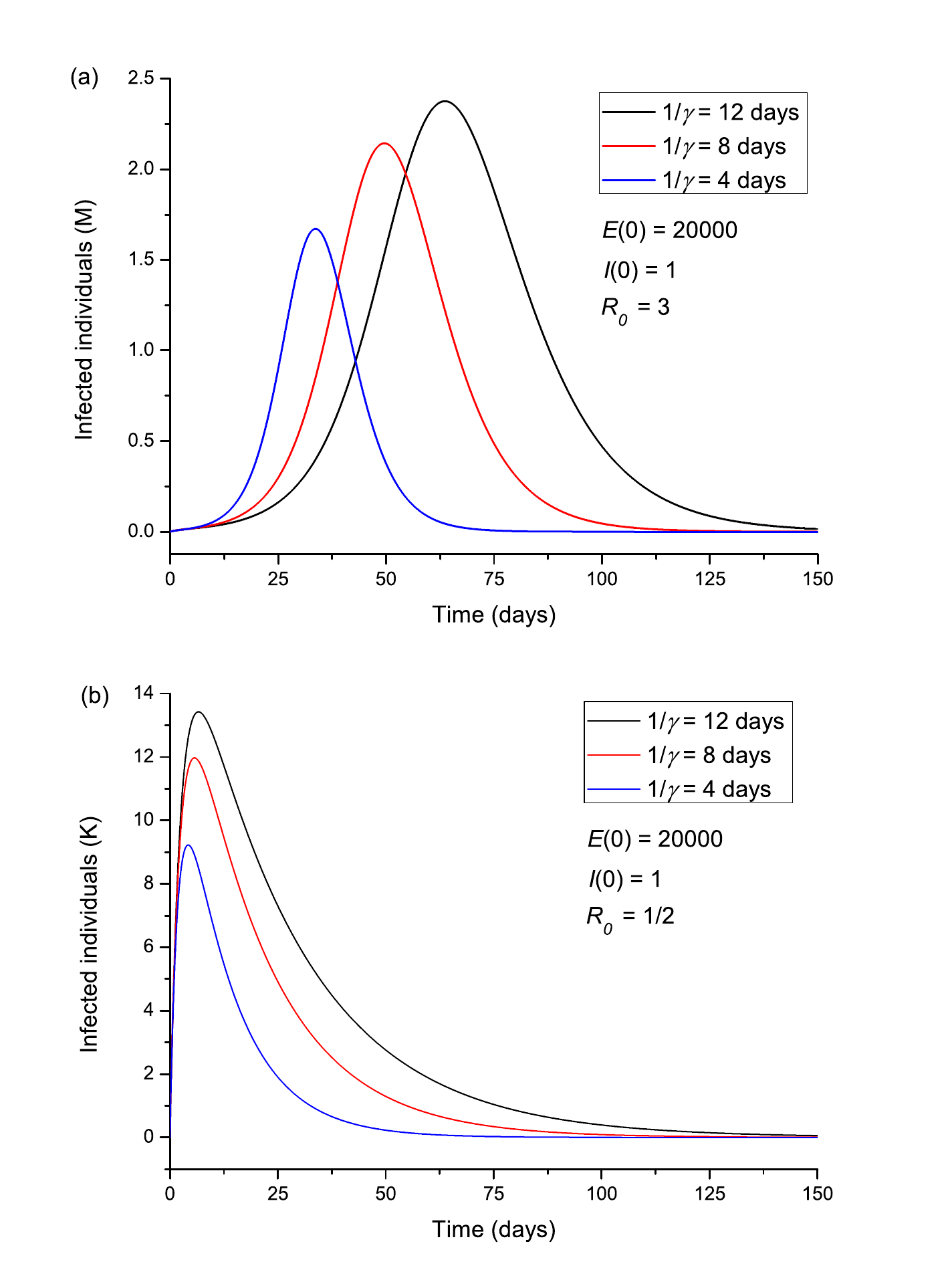}
\caption{Infected individuals for different values of the infectious period $\gamma^{-1}$, 
corresponding to $R_0$ greater (a) and smaller (b) than 1. 
}
\vspace{1cm}
\end{figure}

\begin{figure}
\includegraphics[scale=0.7]{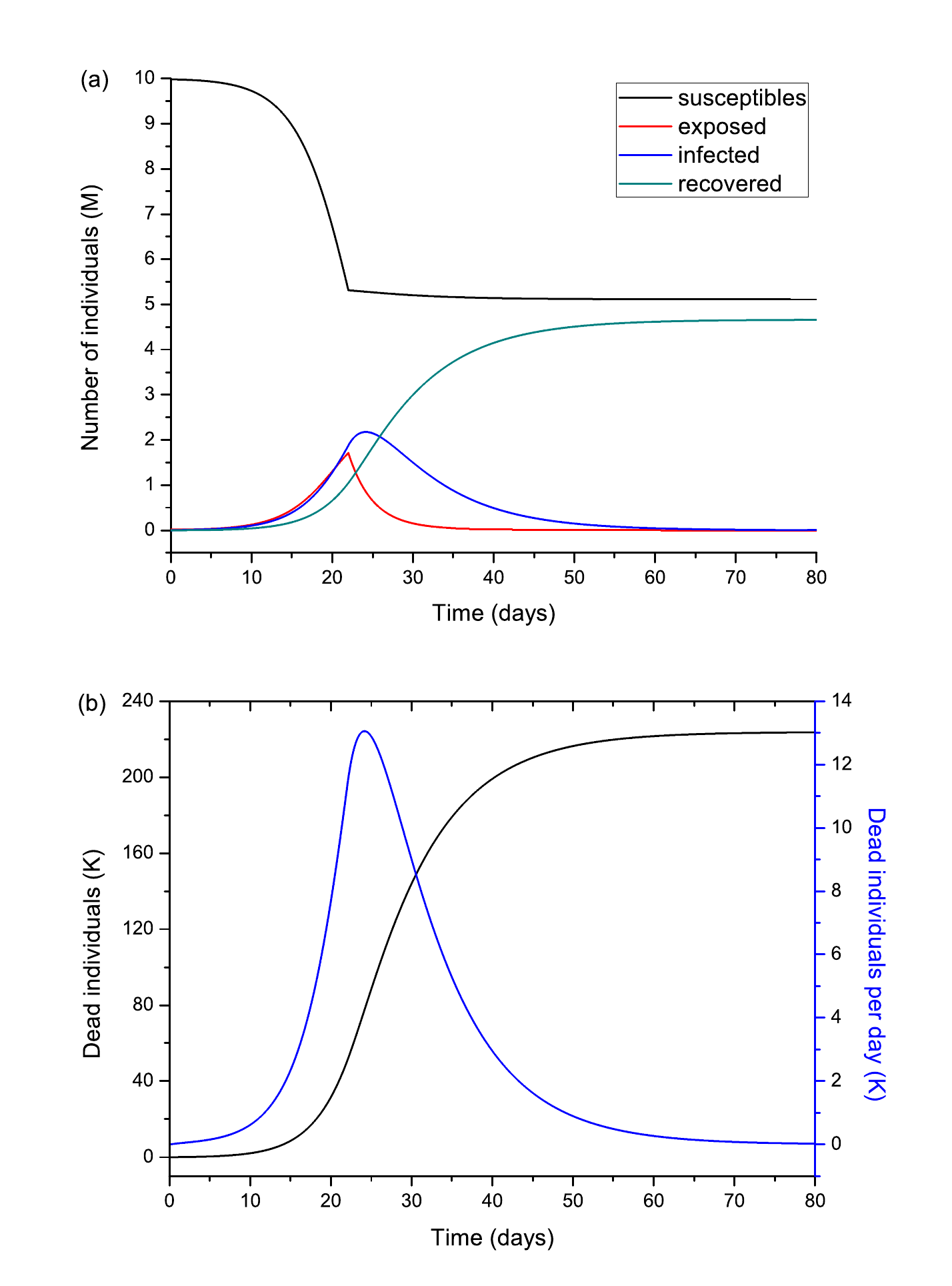}
\caption{Same as Figure 2, but modifying $R_0$ from 5.72 to 0.1 at day 22. 
}
\vspace{1cm}
\end{figure}

\begin{figure}
\includegraphics[scale=0.35]{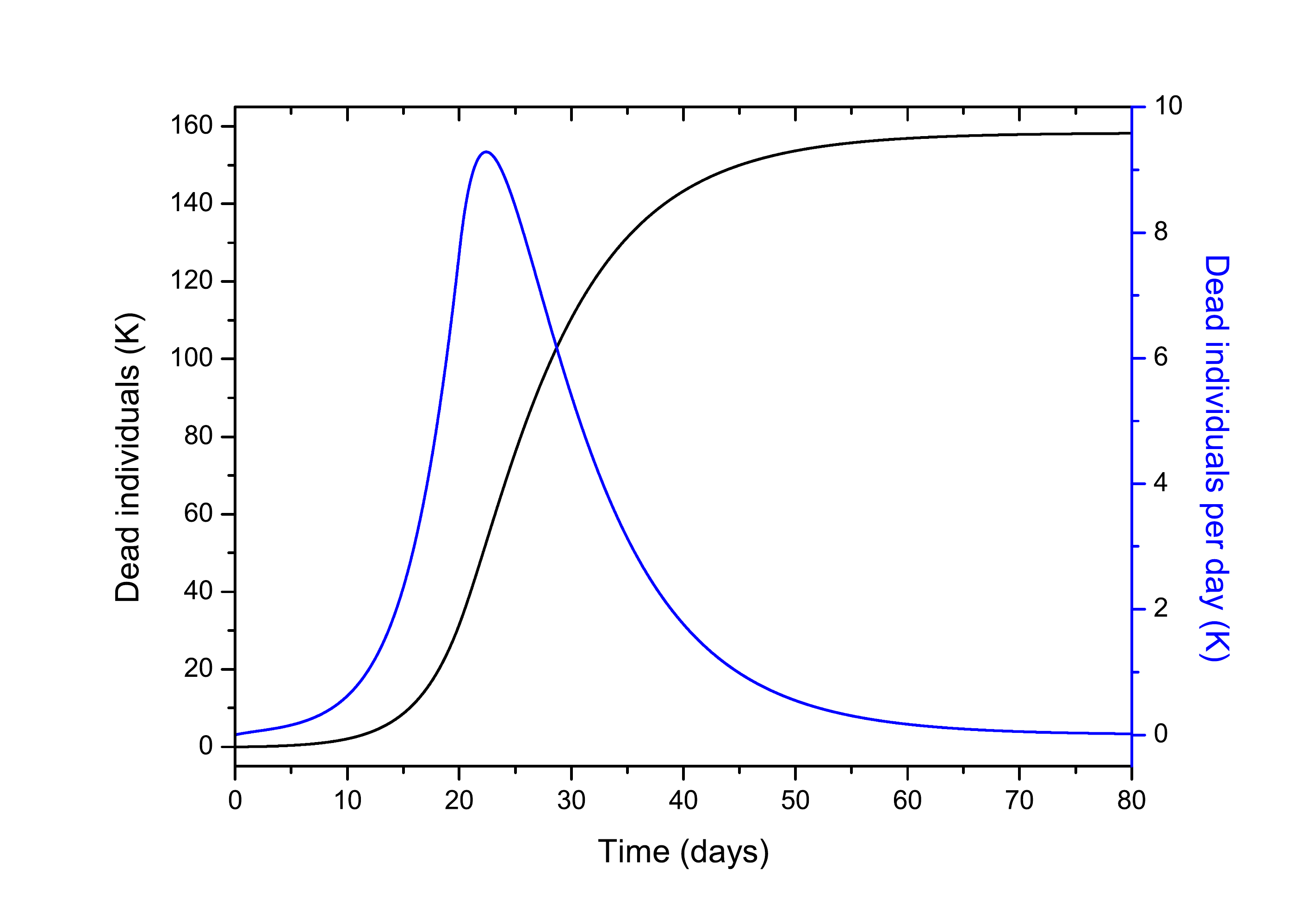}
\caption{Same as Figure 8b, but starting the isolation two days before. 
}
\vspace{1cm}
\end{figure}

\begin{figure}
\includegraphics[scale=0.7]{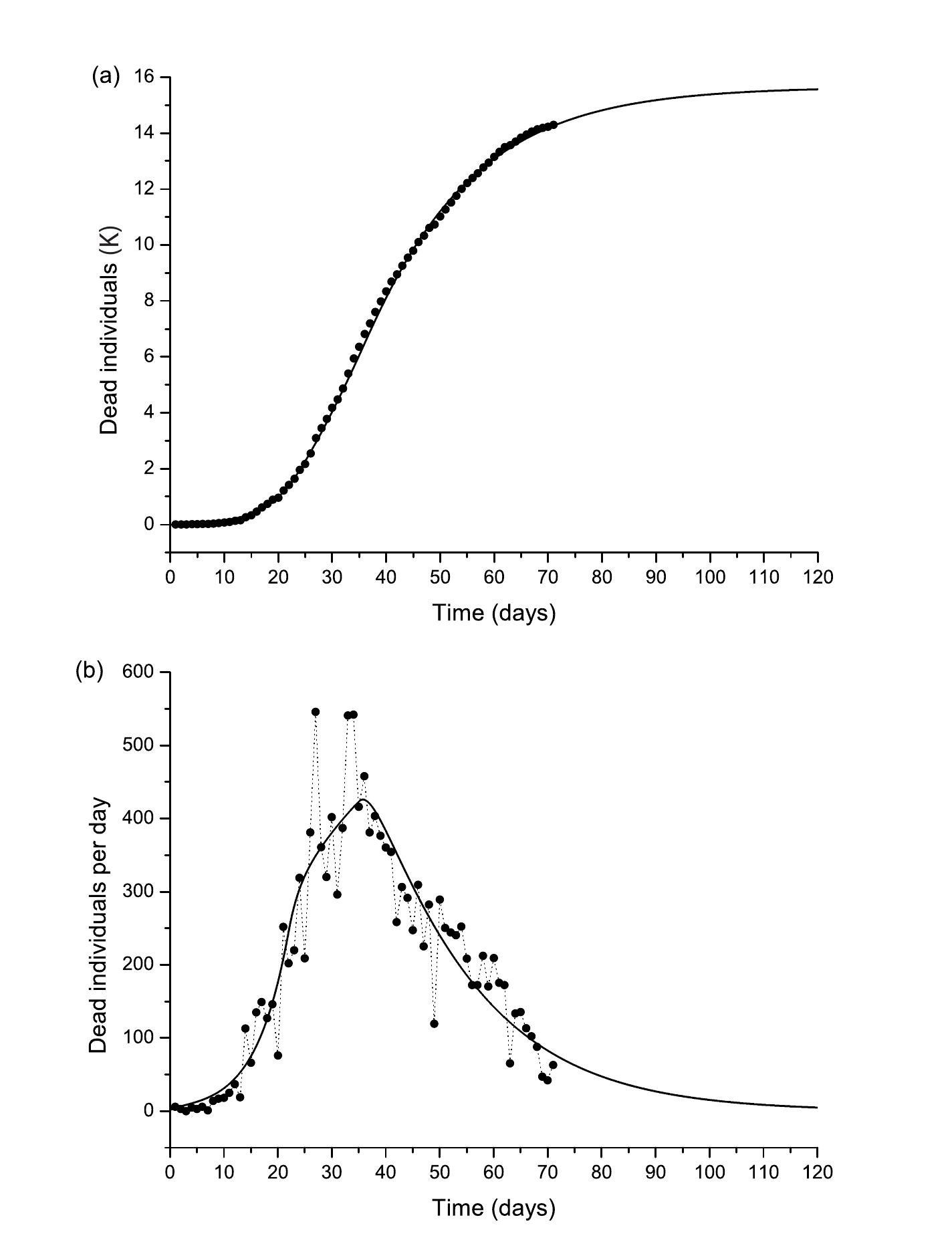}
\caption{
The Lombardy case history. Dead individuals (a) and number of deaths per day (b), where the black  dots represent the data. The solid line corresponds to Case 1 in Table 1. The 
peak can be observed at day 37 (March 31). 
}
\vspace{1cm}
\end{figure}

\begin{figure}
\includegraphics[scale=0.7]{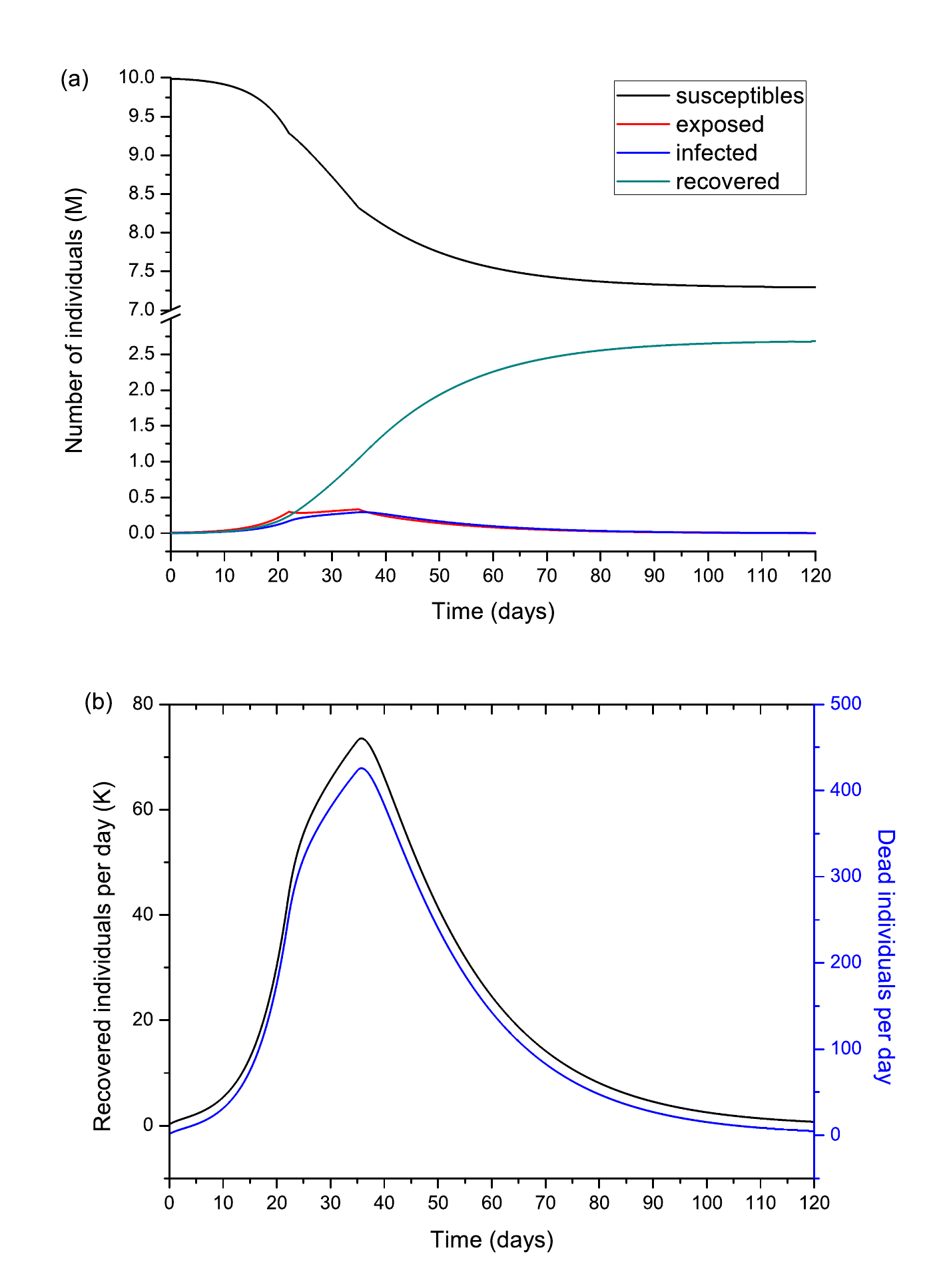}
\caption{Number of humans in the different classes (millions) (a) and recovered individuals per day ($\dot R$) compared to the deaths per day (b) for the case shown in Figure 10. Note that $\dot R$ is given in thousands.}
\vspace{1cm}
\end{figure}

\begin{figure}
\includegraphics[scale=0.7]{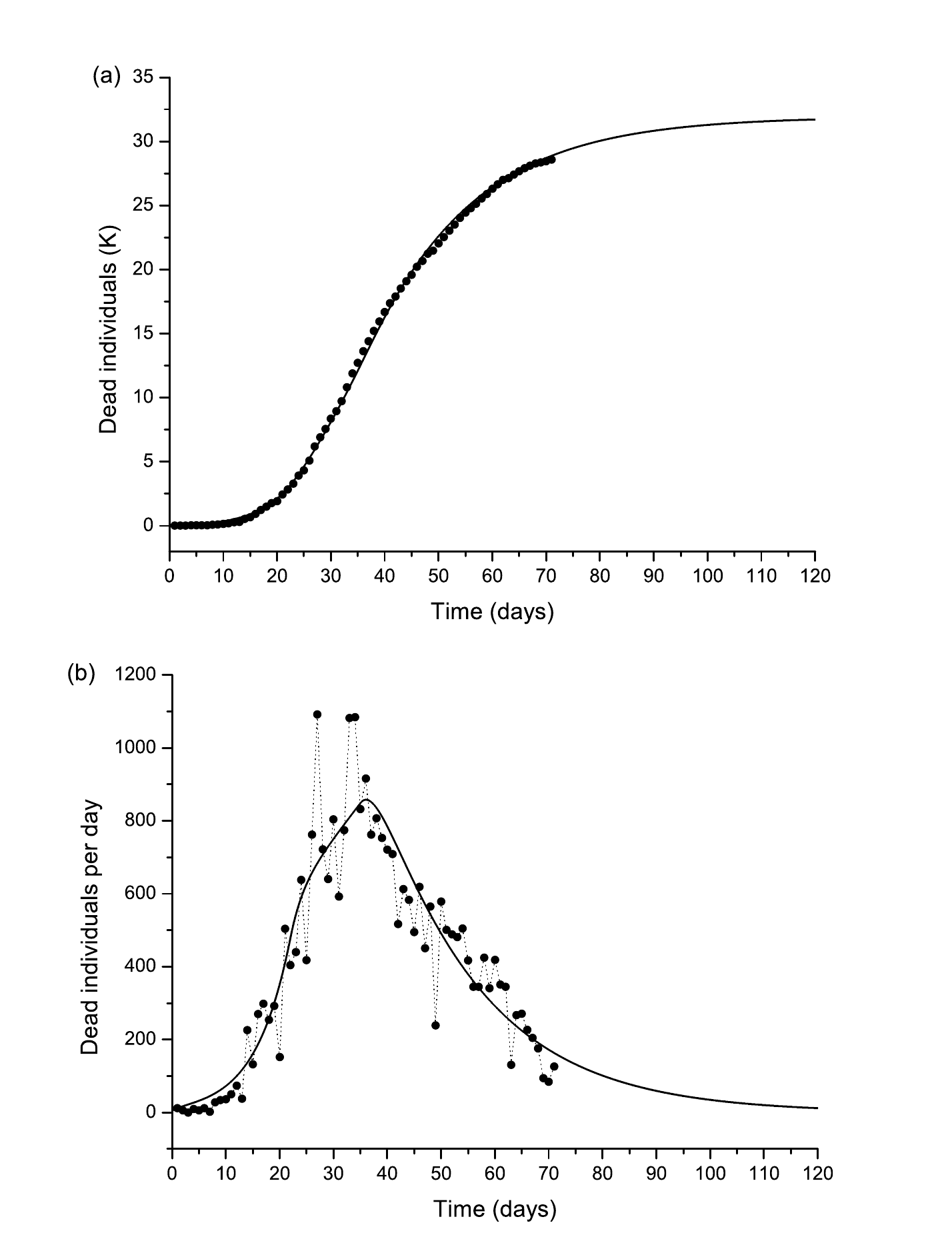}
\caption{Same as Figure 10, but with twice the number of casualties. 
The solid line corresponds to Case 2 in Table 1.}
\vspace{1cm}
\end{figure}

\begin{figure}
\includegraphics[scale=0.7]{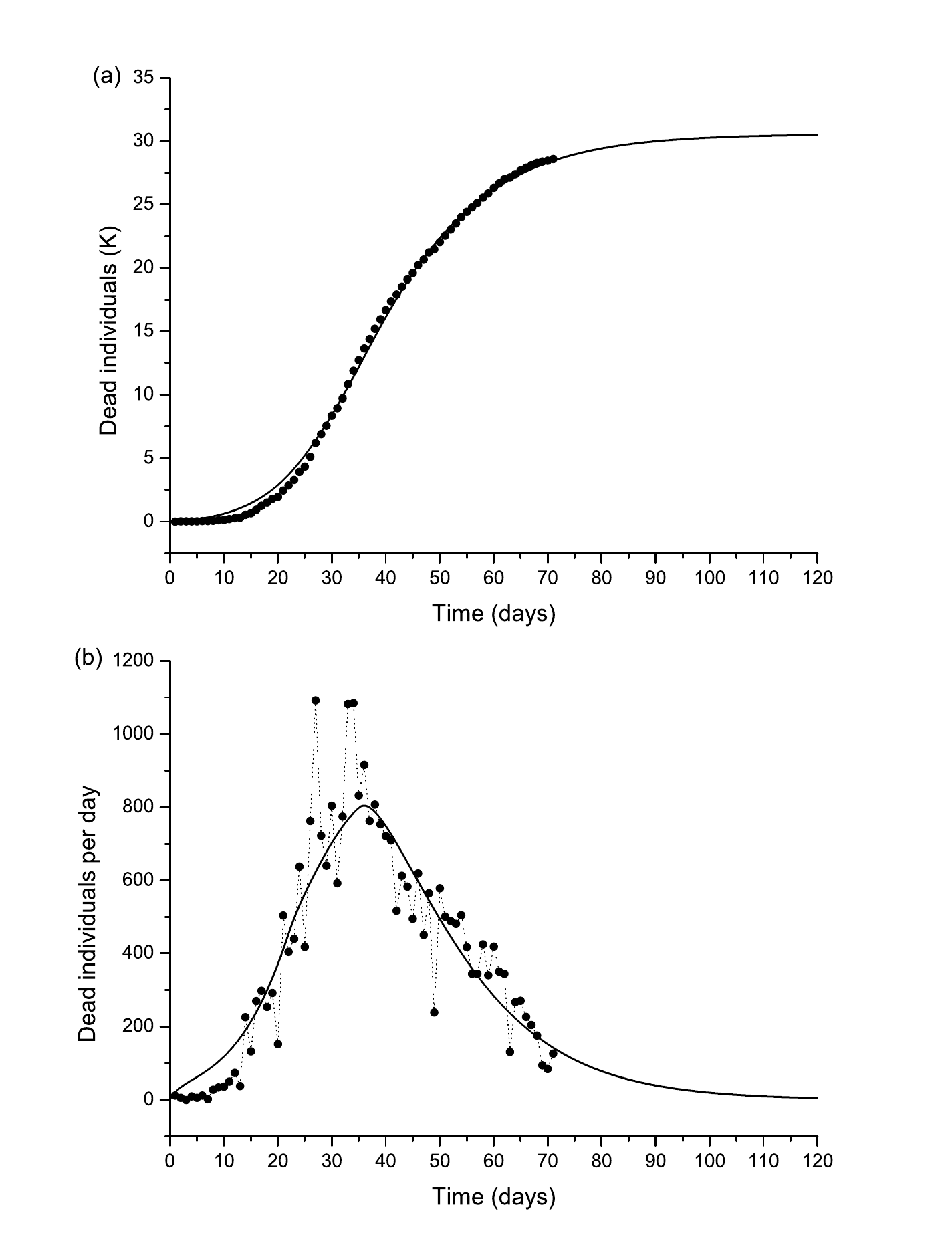}
\caption{Same as Figure 10, but with twice the number of casualties. 
The solid line corresponds to Case 3 in Table 1.}
\vspace{1cm}
\end{figure}

\begin{figure}
\includegraphics[scale=0.35]{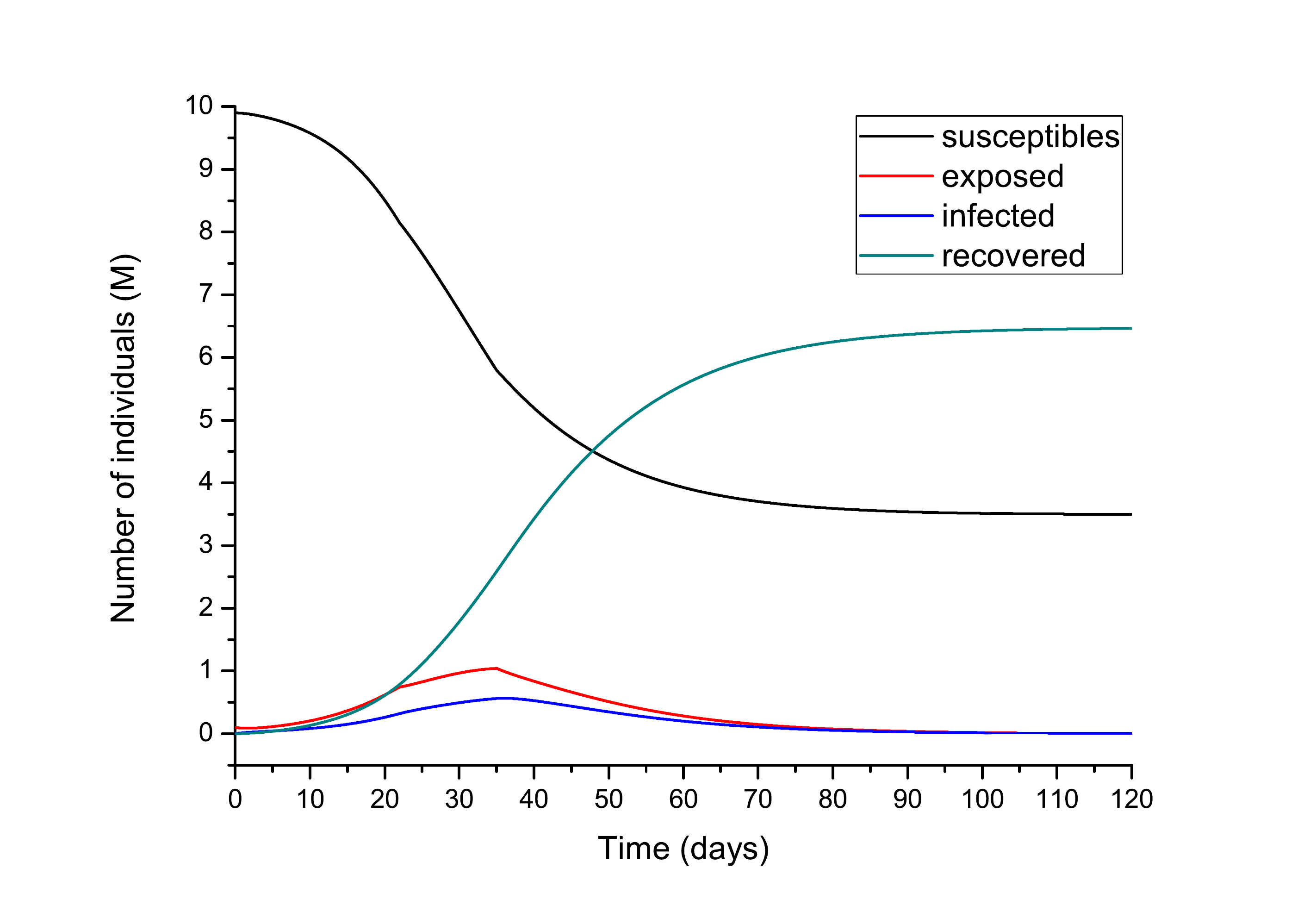}
\caption{Number of humans in the different classes (millions) for the case shown in Figure 13.}
\vspace{1cm}
\end{figure}

\end{document}